\documentclass[aip,amsmath,amssymb,print,superscriptaddress]{revtex4-1}
\usepackage{graphicx}
\usepackage{dcolumn}
\usepackage[colorlinks,linkcolor=blue,anchorcolor=blue,citecolor=blue]{hyperref}
\usepackage{bm}
\begin{document}

\title{\large Multidimensional effects on proton acceleration using high-power intense laser pulses}

\author{K. D. Xiao}
\affiliation{Center for Applied Physics and Technology, HEDPS, and School of Physics, Peking University, Beijing 100871, People's Republic of China}
\author{C. T. Zhou}
\email[]{Electronic mail: zcangtao@iapcm.ac.cn}
\affiliation{Center for Applied Physics and Technology, HEDPS, and School of Physics, Peking University, Beijing 100871, People's Republic of China}
\affiliation{Institute of Applied Physics and Computational Mathematics, Beijing 100094, People's Republic of China}
\affiliation{
College of New Energy and New Materials, Shenzhen Technology University, Shenzhen 518118, People's Republic of China}
\author{K. Jiang}
\affiliation{Graduate School, China Academy of Engineering Physics, P.O. Box 2101, Beijing 100088, People's Republic of China}
\author{Y. C. Yang}
\affiliation{Center for Applied Physics and Technology, HEDPS, and School of Physics, Peking University, Beijing 100871, People's Republic of China}
\author{R. Li}
\affiliation{Center for Applied Physics and Technology, HEDPS, and School of Physics, Peking University, Beijing 100871, People's Republic of China}
\author{H. Zhang}
\affiliation{Institute of Applied Physics and Computational Mathematics, Beijing 100094, People's Republic of China}
\author{B. Qiao}
\affiliation{Center for Applied Physics and Technology, HEDPS, and School of Physics, Peking University, Beijing 100871, People's Republic of China}
\author{T. W. Huang}
\affiliation{College of Optoelectronic Engineering, Shenzhen University, Shenzhen 518060, People's Republic of China}
\author{J. M. Cao}
\affiliation{
College of New Energy and New Materials, Shenzhen Technology University, Shenzhen 518118, People's Republic of China}
\author{T. X. Cai}
\affiliation{
College of New Energy and New Materials, Shenzhen Technology University, Shenzhen 518118, People's Republic of China}
\author{M. Y. Yu}
\affiliation{
College of New Energy and New Materials, Shenzhen Technology University, Shenzhen 518118, People's Republic of China}
\author{S. C. Ruan}
\affiliation{
College of New Energy and New Materials, Shenzhen Technology University, Shenzhen 518118, People's Republic of China}
\affiliation{College of Optoelectronic Engineering, Shenzhen University, Shenzhen 518060,  People's Republic of China}
\author{X. T. He}
\affiliation{Center for Applied Physics and Technology, HEDPS, and School of Physics, Peking University, Beijing 100871, People's Republic of China}
\affiliation{Institute of Applied Physics and Computational Mathematics, Beijing 100094, People's Republic of China}

\date{\today}

\begin{abstract}
Dimensional effects in particle-in-cell (PIC) simulation of target normal sheath acceleration (TNSA) of protons are considered. As the spatial divergence of the laser-accelerated hot sheath electrons and the resulting space-charge electric field on the target backside depend on the spatial dimension, the maximum energy of the accelerated protons obtained from three-dimensional (3D) simulations is usually much less that from two-dimensional (2D) simulations. By closely examining the TNSA of protons in 2D and 3D PIC simulations, we deduce an empirical ratio between the maximum proton energies obtained from the 2D and 3D simulations. This ratio may be useful for estimating the maximum proton energy in realistic (3D) TNSA from the results of the corresponding 2D simulation. It is also shown that the scaling law also applies to TNSA from structured targets.
\end{abstract}

\pacs{52.38.Kd, 52.65.Rr, 52.50.Jm, 29.25.-t}

\maketitle

\section{Introduction}

Laser driven proton acceleration can produce proton beams of high
energy and low divergence, as well as large proton
number.\cite{macchi2013,daido2012,robinson-rpa,zhou-pro1,zhangwl}
Such high-quality energetic proton beams are useful in ultrafast
radiography, tumor therapy, inertial confinement fusion,
etc.\cite{binjh,abicht,xthe,Borghesi2006} The target normal sheath
acceleration (TNSA) scheme is one of the most widely investigated
mechanisms of proton
acceleration.\cite{macchi2013,daido2012,Passonirev,wilks2001,mora}
In TNSA, the intense laser irradiating a thin solid target
generates, heats, and accelerates the electrons on its front
surface. The hot electrons can easily penetrate through the target
and create a huge charge-separation electric field behind its rear
surface, where protons can be accelerated by this electric field to
a few or tens of MeV.\cite{Wagner,gaillard2011,Snavely,fuchs,robson}

Like in many experiments involving complex phenomena, in proton
acceleration it is difficult to scan all the laser and target
parameters due to the high operational cost and limited laser shots.
With rapid development of computational techniques, parallelized
computer simulations are useful for predicting and/or verifying
experimental results and as guide for optimizing target design.
However, full-scale three-dimensional (3D) particle-in-cell (PIC)
simulations are at present still impractical if large regions and
long times of interactions are involved, and two-dimensional (2D) simulations are
often used instead. However, it has been found that the maximum
proton energy (MPE) from 2D PIC simulations of
TNSA is consistently overestimated compared with that from the
3D simulations and the experiments.\cite{sgattoni,Humieres,liujl,blanco}
It is thus of interest to see if there exists a relation between the
TNSA MPEs obtained from 2D and 3D simulations.

In this paper, we perform 2D and 3D PIC simulations of TNSA of
protons under different conditions. By closely examining the
results, we found a ratio of the MPEs from the 2D and 3D
simulations. The empirical ratio is justified by a simple
theoretical model and is consistent with that obtained from
comparing the results from existing 2D simulations with the
relevant experiments. Validity of this ratio for TNSA with
micro-structured targets is also discussed.

The paper is arranged as follows. In Sec. II we compare the 2D and
3D results obtained from PIC simulations of femtosecond laser-driven
TNSA proton acceleration. In Sec. III a model for the ratio of the
2D and 3D MPEs is introduced.
In Sec. IV, the model is applied to picosecond-laser driven
proton acceleration. Sec. V considers the dimensional effects on
TNSA using micro-structured targets. Sec. VI summarizes our
results.

\section{Proton acceleration driven by femtosecond laser pulses}

\begin{figure}
\includegraphics[width=8.5cm]{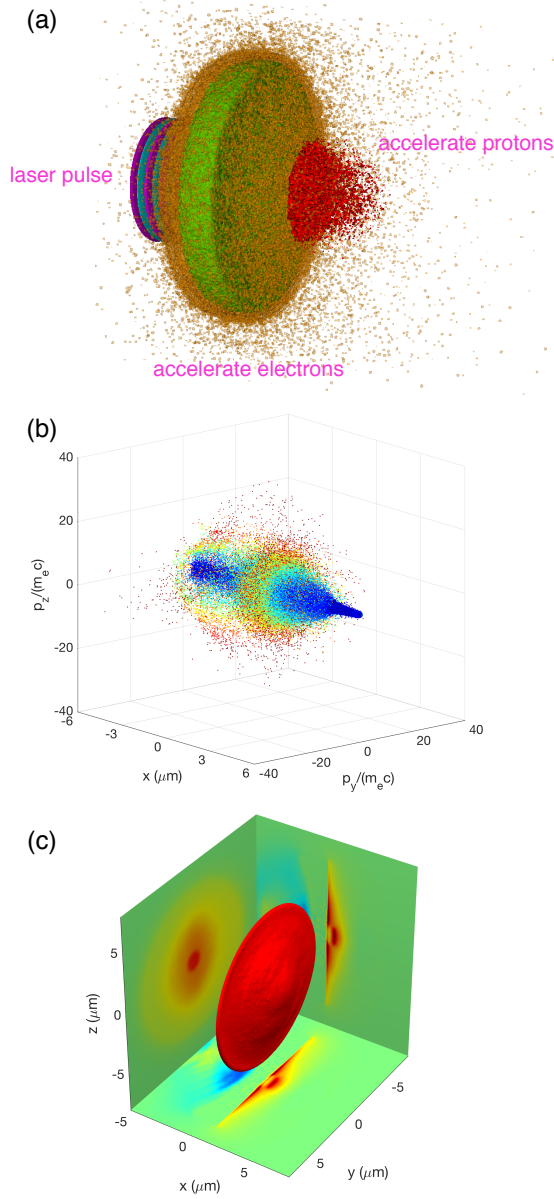}
\caption{\label{fig0} (Color online) (a) Scheme for target normal
sheath acceleration (TNSA) of protons. (b) Distribution of the
transverse momentum ($p_y$ and $p_z$) of the accelerated electrons
along the laser ($x$) direction. (c) Averaged sheath electric field
$E_x$.}
\end{figure}

The scheme for TNSA of protons is illustrated in Fig. \ref{fig0}(a). The
target is assumed to be pre-ionized. A Gaussian laser pulse
irradiates the foil target and the affected electrons on the target
front are accelerated by the laser ponderomotive force. These hot
electrons can transit through the foil target, so that an intense
charge-separation electric field is created behind the foil's rear
surface [see Figs. \ref{fig0}(b) and (c)]. The protons in a dot
source attached to the latter are thereby accelerated by the intense
charge-separation field. As can be seen in Figs. \ref{fig0}(b) and
(c), the hot electrons and the sheath electric field behind the
target are spatially divergent. Since the divergence is dimension
dependent, there can be a difference in the results of the 2D and 3D
PIC simulations of TNSA of protons.

To characterize the dimensional effects of TNSA of protons, two and
three dimensional PIC simulations are carried out using the
\textsc{epoch2d} and \textsc{epoch3d} PIC codes,
respectively.\cite{arber} The flat foil target is composed of overdense
copper plasma with electron number density $n_0=40n_c$, and particle
collisions are neglected in the PIC simulations. (We have also
carried out simulations for other electron densities and for both
with and without particle collisions, but their effects on the
resulting MPEs turn out to be small compared with that of the
dimension.) The mass and charge of the copper ion are 63.5 and $+2$,
respectively. The initial temperatures of the ions and electrons are
$T_i=170\ \rm{eV}$ and $T_e=1\ \rm{keV}$, respectively. The
thickness and width of the target are $1\ \mu\rm{m}$ and $12\
\rm{\mu m}$, respectively. A small-scale preplasma with the profile
$n_e=n_0\exp(x/l)$, where $l=0.3\ \rm{\mu m}$, is placed in front of
the target. A small proton dot of diameter $1\ \rm{\mu m}$ and
thickness $0.5\ \rm{\mu m}$ providing the proton source is attached
to the target backside.\cite{schwoerer,robinson,gibbon,pukhov-prl} A
$y$-polarized Gaussian laser pulse with intensity $1\times10^{21}\
\rm{W/cm}^2$ and wavelength $800\ \rm{nm}$ enters from the left
boundary of the simulation box. The laser spot radius is $3\
\mu\rm{m}$ and the pulse duration is $20\ \rm{fs}$. 
In the 2D simulations, the simulation box is $35\ \mathrm{\mu m}$
and $20\ \mu\mathrm{m}$ in the $x$ and $y$ directions 
with $3279$ and $1987$ spatial grids, respectively. There are 50
electrons and 25 ions in each target cell. In the 3D simulations,
the simulation box is $35\ \mathrm{\mu m}$ and $20\ \mathrm{\mu m}$
and $20\ \mathrm{\mu m}$ in the $x, y,$ and $z$ directions with 
$1640, 994,$ and $994$ spatial grids, respectively. There are 4
electrons and 2 ions in each target cell. The corresponding grid
length is 0.5 skin depth in the 2D and 1 skin depth in the 3D
simulations. The proton dot in the 2D simulations is resolved by
$47$ spatial grids in the $x$ direction and $99$ spatial grids in
the $y$ direction. There are $500$ proton macroparticles in each
cell. The proton dot in the 3D simulations is resolved by $23$
spatial grids in the $x$ direction and $49$ spatial grids in the
$y$- and $z$-directions. There are $40$ proton macroparticles in
each cell. Periodic boundary conditions are used in the transverse
directions and open boundaries are used in the longitudinal
directions. The laser enters from the left boundary of the box.

\begin{figure*}
\includegraphics[width=15cm]{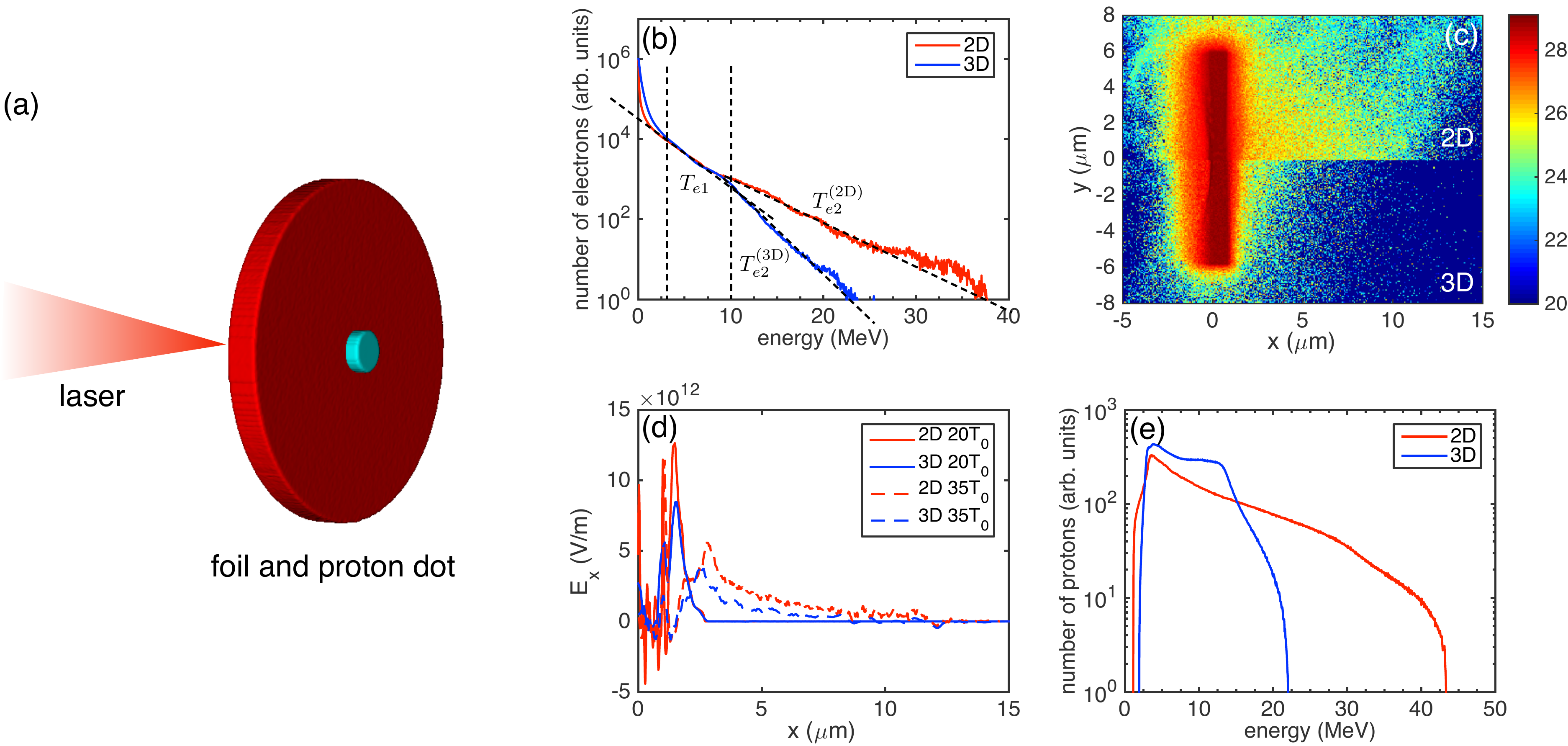}
\caption{\label{fig1} (Color online) Simulation results for the flat
foil target. (a) Target setup. (b) Electron spectrum at $t=30{T_0}$.
(c) Electron number density at $t=100{T_0}$. (d) Profile of the
electric field $E_x$ at $t=20{T_0}$ and $35{T_0}$, respectively. (e)
Proton spectrum at $t=100{T_0}$. }
\end{figure*}

\begin{figure}
\includegraphics[width=9cm]{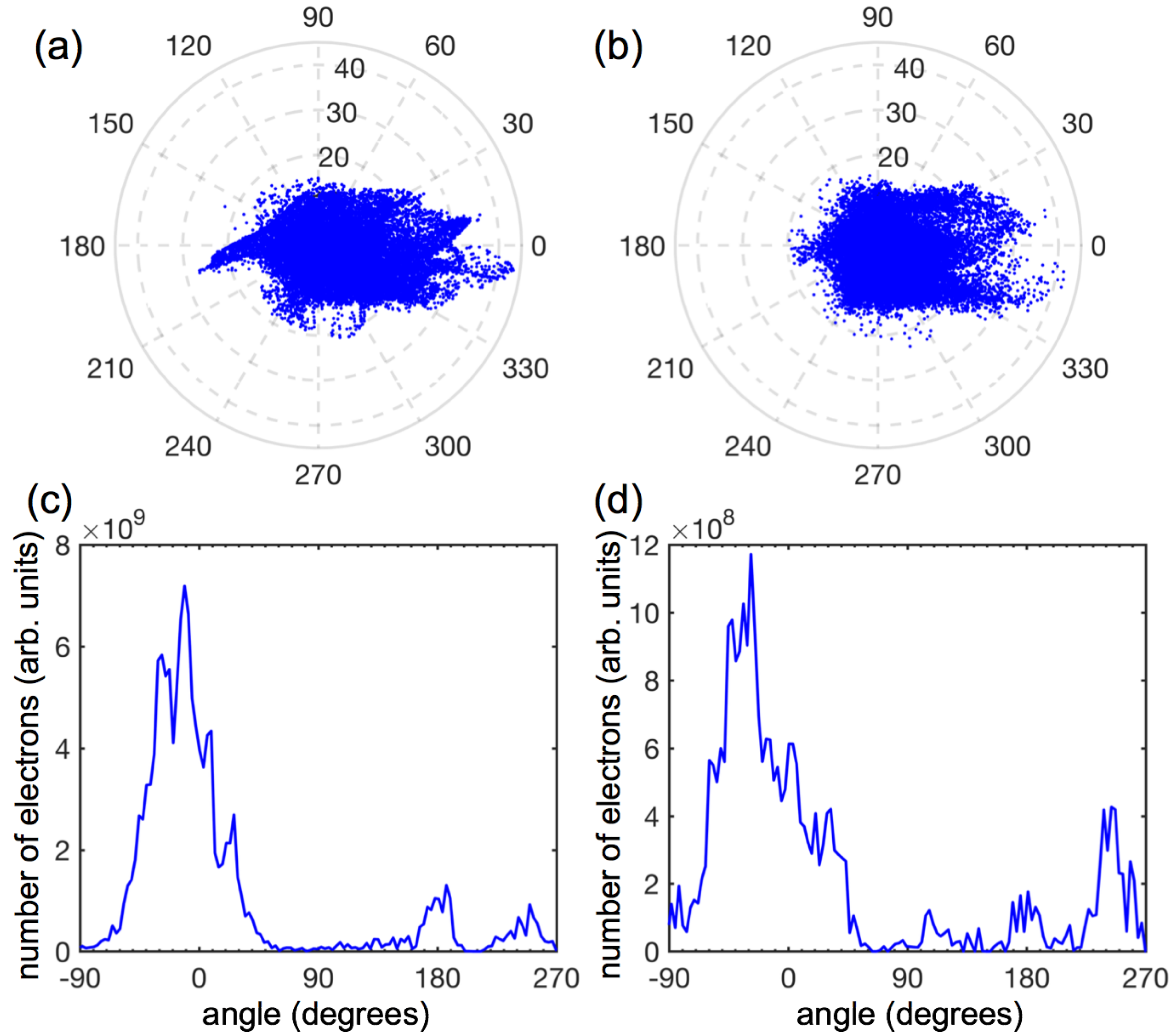}
\caption{\label{fig_div} (Color online.) (a) and (b) Distribution of 
electron divergence in the ($\epsilon_e, \theta_{xy}$) 
polar plane from the 2D and 3D simulations, respectively, at
$t=10{T_0}$. Here, $\epsilon_e=(\gamma-1)m_ec^2$ and $\theta_{xy}=\arctan(p_y/p_x)$. 
(c) and (d) The corresponding angular distribution of the number of electrons with energies greater than $10\ \rm{MeV}$.}
\end{figure}

Figures \ref{fig1} and \ref{fig_div} show the 2D and 3D simulation
results. The electron energy spectrums are compared in Fig.
\ref{fig1}(b). The electron number densities are shown in Fig. \ref{fig1}(c). 
As can be seen in Fig. \ref{fig_div}, the divergence of the hot
electrons is larger in 3D. The larger divergence
in the 3D case reduces the electron recirculation rate inside
the target, thereby weakening the electron heating. Fig. \ref{fig1}(b)
shows that electrons with energies greater than $10\ \rm{MeV}$ as
well as the maximum electron energy are lower in 3D than in 2D.
Also, as shown in Figs. \ref{fig1}(b) and \ref{fig1}(c), the
number of electrons with energies greater than $10\ \rm{MeV}$ as
well as the electron number density, in the 3D case is smaller. 
Fig. \ref{fig1}(c) shows that due to the additional degree of
freedom of the affected electrons, the laser hole boring depth in 3D
is deeper than in 2D. According to the plasma expansion model of
Mora,\cite{mora} the strength of the sheath field can be estimated
by $E_{\rm{sheath}}=2(4\pi
n_eT_e)^{1/2}/(2e+\omega_{pi}^2t^2)^{1/2}$, where $e=2.71828$ is the
Euler number and $\omega_{pi}=(4\pi n_ee^2/m_p)^{1/2}$ is the ion
plasma frequency. Thus, the strength of the sheath field is
$\propto(n_eT_e)^{1/2}$.
Since both the electron number density and temperature are lower in
3D, the sheath electric field $E_x$ at the target backside is
weaker than that in 2D. In fact, we have [see Fig. \ref{fig1} (d)]
$E_x^{\rm{(2D)}}=1.3\times10^{13}\ \rm{V/m}$ and
$E_x^{\rm{(3D)}}=8\times10^{12}\ \rm{V/m}$ at $t=20{T_0}$, and
$E_x^{\rm{(2D)}}=6\times10^{12}\ \rm{V/m}$ and
$E_x^{\rm{(3D)}}=4\times10^{12}\ \rm{V/m}$ at $t=35{T_0}$.
As a result, the proton energy from 3D simulation is lower than that from 2D. 
As shown in Fig. \ref{fig1}(f) for the proton energy
spectrums, the MPE is $22\ \rm{MeV}$ in 3D and $43\
\rm{MeV}$ in 2D.

\section{Qualitative model on proton energy difference}

As mentioned, it is of practical interest to find a relation
between the MPEs obtained from the 2D and the more realistic 3D
simulations.
In this section, we investigate the dependence of the energy ratio
on the laser and target parameters and give a qualitative model
for this ratio.

\begin{table*} \scriptsize
\caption{Comparison of maximum proton energies from 2D and 3D
simulations for different laser and target parameters. The energy
ratio given by the qualitative model is 2.17 for lasers with spot
radius $3\ \rm{\mu m}$, and 2.51 for lasers with spot radius
$4\ \rm{\mu m}$.}
\begin{center}
\begin{tabular}{cccccc}
\hline\hline
laser intensity & spot and duration & target thickness & proton energy (3D) & proton energy (2D) & energy ratio (2D/3D)\\
\hline
$5\times10^{19}\ \rm{W/cm^2}$ & $3\ \rm{\mu m}$, $20\ \rm{fs}$ & $1\ \rm{\mu m}$ & $4\ \rm{MeV}$ & $8\ \rm{MeV}$ &  2.00\\
$5\times10^{19}\ \rm{W/cm^2}$ & $3\ \rm{\mu m}$, $100\ \rm{fs}$ & $1\ \rm{\mu m}$ & $10\ \rm{MeV}$ & $20\ \rm{MeV}$ &  2.00\\
$2\times10^{20}\ \rm{W/cm^2}$ & $4\ \rm{\mu m}$, $20\ \rm{fs}$ & $1\ \rm{\mu m}$ & $9\ \rm{MeV}$ & $21\ \rm{MeV}$ & 2.33 \\
$2\times10^{20}\ \rm{W/cm^2}$ & $4\ \rm{\mu m}$, $100\ \rm{fs}$ & $1\ \rm{\mu m}$ & $19\ \rm{MeV}$ & $46\ \rm{MeV}$ & 2.42 \\
$1\times10^{21}\ \rm{W/cm^2}$ & $3\ \rm{\mu m}$, $20\ \rm{fs}$ & $1\ \rm{\mu m}$ & $22\ \rm{MeV}$ & $43\ \rm{MeV}$ & 1.95 \\
$1\times10^{21}\ \rm{W/cm^2}$ & $3\ \rm{\mu m}$, $20\ \rm{fs}$ & $3\ \rm{\mu m}$ & $18\ \rm{MeV}$ & $37\ \rm{MeV}$ & 2.05 \\
$1\times10^{21}\ \rm{W/cm^2}$ & $3\ \rm{\mu m}$, $20\ \rm{fs}$ & $6\ \rm{\mu m}$ & $16\ \rm{MeV}$ & $33\ \rm{MeV}$ & 2.06 \\
$1\times10^{21}\ \rm{W/cm^2}$ & $3\ \rm{\mu m}$, $100\ \rm{fs}$ & $1\ \rm{\mu m}$ & $41\ \rm{MeV}$ & $88\ \rm{MeV}$ & 2.15 \\
\hline\hline
\end{tabular}
\end{center}
\label{table1}
\end{table*}

The results are shown in Table \ref{table1}. The simulation setups
in 2D and 3D are the same as that in Sec. II. The laser intensities
are $5\times10^{19}\ \rm{W/cm^2}$, $2\times10^{20}\ \rm{W/cm^2}$,
and $1\times10^{21}\ \rm{W/cm^2}$. For each intensity, two pulse
durations, namely $20\ \rm{fs}$ and $100\ \rm{fs}$, are
investigated. It is found that the ratio between the resulting 2D
and 3D MPEs is from $2$ to $2.5$ for laser spot radii from $3$ to
$4\ \rm{\mu m}$, respectively. That is, within the considered domain, the
dependence on the laser intensity is small. Moreover, the differences
in the 2D and 3D MPEs for laser durations $20\ \rm{fs}$ and $100\ \rm{fs}$
are negligible. 
Simulations are also carried out for target thicknesses $3\ \rm{\mu
m}$ and $6\ \rm{\mu
m}$, laser intensity $1\times10^{21}\ \rm{W/cm^2}$, pulse duration $20\ \rm{fs}$, and spot radius $3 \rm{\mu m}$. The results are
also given in Table \ref{table1}. We see that, compared with that of
the $1\ \rm{\mu m}$ target, the proton energies for the target
thicknesses $3\ \rm{\mu m}$ and $6\ \rm{\mu
m}$ are less, which can be attributed to reduced
electron recirculation inside the thicker target.\cite{Mackinnon}
However, the corresponding 2D to 3D MPE ratios are about
2.05 and 2.06, which are near the value 1.95 for the $1\ \rm{\mu m}$
target.

We now present a simple model for
the 2D to 3D MPE ratio. We assume that the electron distribution
in laser foil interaction is double Maxwellian, as shown in Fig.
\ref{fig1}(b), where
$n(\epsilon)=n_{\mathrm{cold}}(\epsilon)+n_{\mathrm{hot}}(\epsilon)
=\theta_{\mathrm{cold}}\exp(-\epsilon/T_{\mathrm{cold}})
+\theta_{\mathrm{hot}}\exp(-\epsilon/T_{\mathrm{hot}})$. The cold
electrons with energy less than $3\ \rm{MeV}$ contribute little to
the sheath field since they are mostly reflected by the huge surface
potential at the target rear. The hot electrons with energy larger
than $3\ \rm{MeV}$, as shown in Fig. \ref{fig1}(b), can be separated
into two parts. The first part $e1$, which consists of electrons
with energy higher than $3\ \rm{MeV}$ and lower than $10\ \rm{MeV}$,
is almost the same in the 2D and 3D simulations. The second part
$e2$, which consists of electrons with energy greater than $10\
\rm{MeV}$, is quite different in the 2D and 3D simulations. That is,
the hot electrons can also be separated into two groups with
different temperatures:
$n_{\rm{hot}}(\epsilon)=\theta_{e1}\exp(-\epsilon/T_{e1})
+\theta_{e2}\exp(-\epsilon/T_{e2})$. From Fig. \ref{fig1}(b), the
fitted temperature for the $e1$ electrons is the same in 2D and 3D
cases with $T_{e1}^{\rm{(2D)}}\approx T_{e1}^{\rm{(3D)}}\approx2.4\
\rm{MeV}$. The temperature for the $e2$ electrons is
$T_{e2}^{\rm{(2D)}}\approx4.2\ \rm{MeV}$ in the 2D case and
$T_{e2}^{\rm{(3D)}}\approx1.9\ \rm{MeV}$ in the 3D case. The total
number of hot electrons can be estimated from the energy relation
$N_{\rm{total}}\sim\eta E_{\rm{laser}}/\bar\epsilon_e$, where
$N_{\rm{total}}$ is the total number of hot electrons, $\eta$ is the
laser-electron energy conversion efficiency, $E_{\rm{laser}}$ is the
input laser energy, and $\bar\epsilon_e$ is the averaged electron
kinetic energy. From the electron spectrums in Fig. \ref{fig1}(b),
we find by integration over the spectrums that the total numbers of
hot electrons with energy greater than $3\ \rm{MeV}$ is nearly equal
in the 2D and 3D simulations, and the ratio is
$\frac{N_{\rm{total}}^{\rm{(2D)}}}{N_{\rm{total}}^{\rm{(3D)}}}
\approx1.05$. The $e2$ electrons only make up a small fraction of
the total hot electrons. The fraction of the $e2$ electrons is
$\frac{N_{e2}^{\rm{(2D)}}}{N_{\rm{total}}}\approx15\%$ in 2D and
$\frac{N_{e2}^{\rm{(3D)}}}{N_{\rm{total}}}\approx5\%$ in 3D. We can
obtain from the energy spectrums the relation
$\sqrt\frac{N_{e1}^{\rm{(2D)}}T_{e1}^{\rm{(2D)}}
+N_{e2}^{\rm{(2D)}}T_{e2}^{\rm{(2D)}}}
{N_{e1}^{\rm{(3D)}}T_{e1}^{\rm{(3D)}}
+N_{e2}^{\rm{(3D)}}T_{e2}^{\rm{(3D)}}}\approx1.06$, which shows that
the laser-to-hot electron energy conversion efficiency in the 2D and
3D simulations is about the same.

The strength of the sheath electric field at the target rear surface
can be approximated by $E_{\rm{sheath}}\sim T_e/e\lambda_D$, where
$\lambda_D=\sqrt{\frac{\epsilon_0T_e}{n_{e}e^2}}$ is the Debye
length and here $n_e$ is the hot electron density. The ratio of the
sheath fields in 2D and 3D can then be expressed as
\begin{equation} 
R_{\rm{sheath}}=\frac{E_{\rm{sheath}}^{\rm{(2D)}}} {E_{\rm{sheath}}^{\rm{(3D)}}}
\approx\sqrt{\frac{n_{e}^{\rm{(2D)}}
T_e^{\rm{(2D)}}}{n_{e}^{\rm{(3D)}} T_e^{\rm{(3D)}}}}
\approx\sqrt{\frac{\frac{N_{e1}^{\rm{(2D)}}
T_{e1}^{\rm{(2D)}}+N_{e2}^{\rm{(2D)}} T_{e2}^{\rm{(2D)}}}{2\sigma}}
{\frac{N_{e1}^{\rm{(3D)}}T_{e1}^{\rm{(3D)}}
+N_{e2}^{\rm{(3D)}}T_{e2}^{\rm{(3D)}}}{\pi\sigma^2}}}.
\end{equation}
From the calculations given in the above paragraph, we have obtained $\sqrt\frac{N_{e1}^{\rm{(2D)}}T_{e1}^{\rm{(2D)}}
+N_{e2}^{\rm{(2D)}}T_{e2}^{\rm{(2D)}}}
{N_{e1}^{\rm{(3D)}}T_{e1}^{\rm{(3D)}}
+N_{e2}^{\rm{(3D)}}T_{e2}^{\rm{(3D)}}}\approx1$ by using the simulation data. Therefore,
\begin{equation} \label{eq1}
R_{\rm{sheath}}=\frac{E_{\rm{sheath}}^{\rm{(2D)}}} {E_{\rm{sheath}}^{\rm{(3D)}}}
\approx\sqrt{\frac{\pi\sigma}{2}}.
\end{equation}
The protons in the dot source at the target rear surface are
accelerated by the sheath field, and their energy is
$\mathcal{E}_{\rm{proton}}\approx e\int
E_{\rm{sheath}}ds_{\rm{acc}}$, where $s_{\rm{acc}}$ is the proton
acceleration distance. In Eq. \ref{eq1}, the ratio of the sheath fields $R_{\rm{sheath}}$ is independent of the acceleration distance. Thus, the ratio of the 2D and 3D TNSA proton
energies is then approximately
\begin{equation}\label{eq2}
\frac{\mathcal{E}_{\rm{proton}}^{\rm{(2D)}}}
{\mathcal{E}_{\rm{proton}}^{\rm{(3D)}}}\approx
\frac{R_{\rm{sheath}}\int E_{\rm{sheath}}^{\rm{(3D)}}ds_{\rm{acc}}}{\int E_{\rm{sheath}}^{\rm{(3D)}}ds_{\rm{acc}}}\approx
\sqrt{\frac{\pi\sigma}{2}},
\end{equation}
where $\sigma$ is the laser spot radius in units of $\rm{\mu m}$. 
We see that the energy ratio depends on the laser spot radius, and
it is 2.17 for $\sigma = 3\ \rm{\mu m}$, and 2.51 for $\sigma = 4\
\rm{\mu m}$. These estimated energy ratios agree fairly well with
that obtained from the simulations, as can be seen in Table
\ref{table1}.

Our model is based on the assumption that the proton
acceleration distance is the same in 2D and 3D. This is
reasonable because in TNSA, the acceleration is mainly in
the axial ($x$) direction. Moreover, the protons gain energy
within a few Debye lengths, where the sheath field is peaked near
the target-vacuum interface.\cite{Snavely,passoni} After this
effective acceleration region, the sheath field decays rapidly in
all directions because of expansion of the hot electrons. However,
the formula may not be applicable if the laser spot is large compared with the foil thickness. If
we assume that the hot electron transport is ballistic,\cite{coury}
the electron divergence effect can be neglected when the electron
transverse displacement inside the foil is far less than the laser
spot size, or $d\cdot\tan\theta_{xy}/\sigma\ll1$, where $d$ is the target
thickness, and $\theta_{xy}$ is the divergence angle. In this case, the
proton acceleration would be roughly one dimensional, so that the
TNSA proton energies from the 2D and 3D simulations would be
similar. In the simulations of femtosecond laser-foil interactions
in Sec. II, the hot electron (for electrons with energies greater
than $10\ \rm{MeV}$) divergence angles $\theta_{xy}$ in the
($x,y$) plane is about $40^{\circ}$ in 2D simulation and
$50^{\circ}$ in 3D simulation [see Figs. \ref{fig_div}(c) and
\ref{fig_div}(d)]. Straightforward calculations show that
$d\cdot\tan\theta_{xy}/\sigma\approx2/7$ in 2D and
$d\cdot\tan\theta_{xy}/\sigma\approx2/5$ in 3D for target
thickness $1\ \rm{\mu m}$ and laser spot radius $3\ \rm{\mu m}$.
These values suggest that multidimensional effects should be
considered.

\begin{figure}
\includegraphics[width=6.5cm]{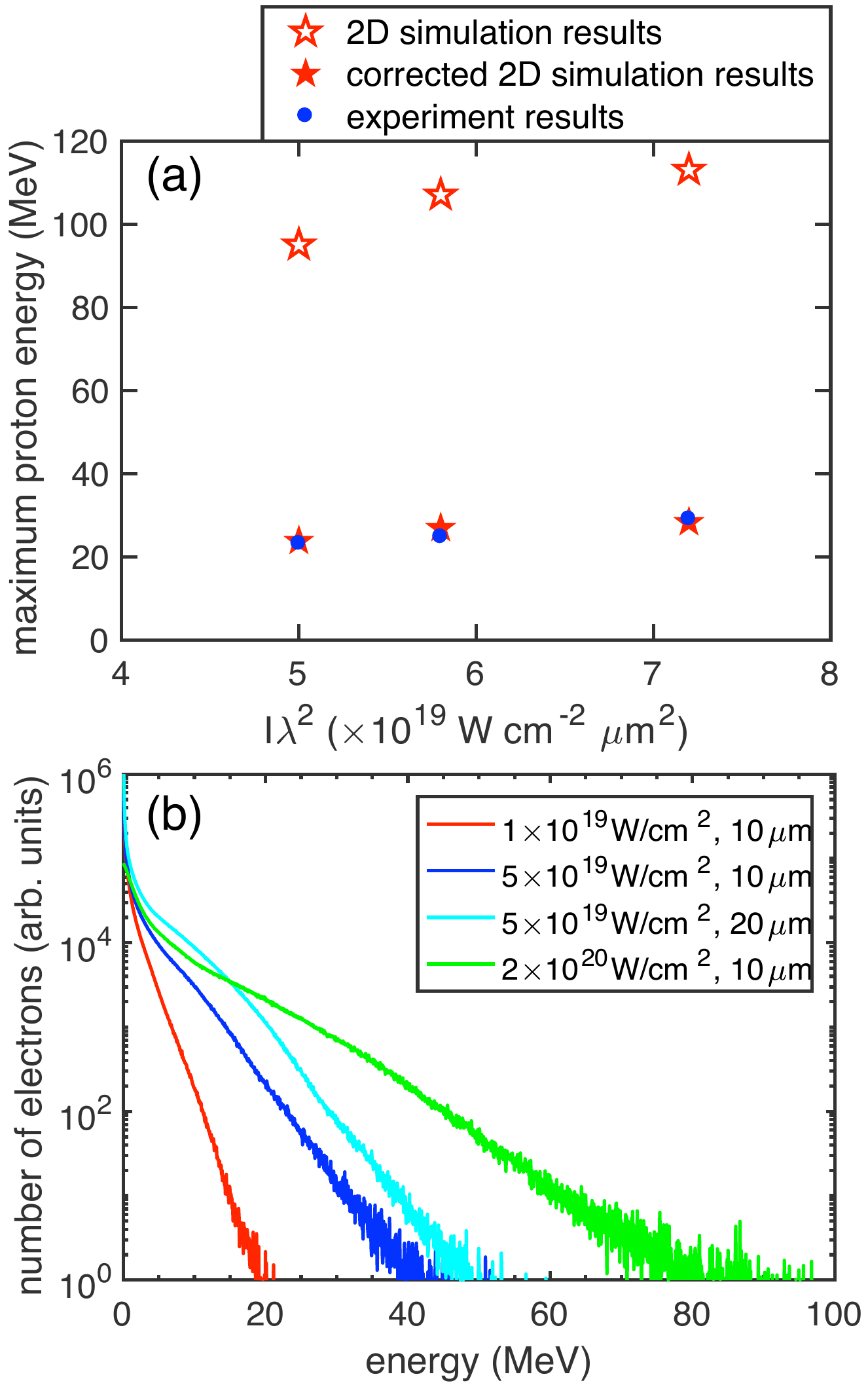}
\caption{\label{fig4} (Color online) (a) Comparison of corrected
simulation results and experiment results of MPE driven by
petawatt-picosecond laser pulses. The blue dots are the experiment
results from the SGII-U laser facility at Shanghai, China. The solid red
stars are the corrected 2D simulation results by dividing the
pristine 2D simulation results (shown in hollow red stars) by the energy ratio 3.96. The
pristine and corrected 2D simulation results are shown in Table
\ref{table2}. (b) Electron spectrums at $t=0.5\ \rm{ps}$ for the
simulated cases. The legend in (b) gives the laser intensity and
preplasma length.}
\end{figure}

\begin{table*} \scriptsize
\caption{Maximum proton energies from petawatt-picosecond laser-foil
interactions.}
\begin{center}
\begin{tabular}{ccccccc}
\hline\hline
laser intensity & spot and duration & preplasma length & target thickness & 2D simulation results & corrected results & experiment results \\
\hline
$1\times10^{19}\ \rm{W/cm^2}$ & $10\ \rm{\mu m}$, $1\ \rm{ps}$ & $10\ \rm{\mu m}$ & $10\ \rm{\mu m}$ & $36\ \rm{MeV}$ & $9\ \rm{MeV}$ & -- \\
$5\times10^{19}\ \rm{W/cm^2}$ & $10\ \rm{\mu m}$, $1\ \rm{ps}$ & $10\ \rm{\mu m}$ & $10\ \rm{\mu m}$ & $82\ \rm{MeV}$ & $21\ \rm{MeV}$ & -- \\
$5\times10^{19}\ \rm{W/cm^2}$ & $10\ \rm{\mu m}$, $1\ \rm{ps}$ & $20\ \rm{\mu m}$ & $10\ \rm{\mu m}$ & $85\ \rm{MeV}$ & $21\ \rm{MeV}$ & -- \\
$5\times10^{19}\ \rm{W/cm^2}$ & $10\ \rm{\mu m}$, $1\ \rm{ps}$ & $80\ \rm{\mu m}$ & $10\ \rm{\mu m}$ & $95\ \rm{MeV}$ & $23.8\ \rm{MeV}$ & $23.2\ \rm{MeV}$ \\
$5.8\times10^{19}\ \rm{W/cm^2}$ & $10\ \rm{\mu m}$, $1\ \rm{ps}$ & $80\ \rm{\mu m}$ & $15\ \rm{\mu m}$ & $107\ \rm{MeV}$ & $26.9\ \rm{MeV}$ & $24.8\ \rm{MeV}$ \\
$7.2\times10^{19}\ \rm{W/cm^2}$ & $10\ \rm{\mu m}$, $1\ \rm{ps}$ & $80\ \rm{\mu m}$ & $15\ \rm{\mu m}$ & $113\ \rm{MeV}$ & $28.3\ \rm{MeV}$ & $29.1\ \rm{MeV}$ \\
$2\times10^{20}\ \rm{W/cm^2}$ & $10\ \rm{\mu m}$, $1\ \rm{ps}$ & $10\ \rm{\mu m}$ & $10\ \rm{\mu m}$ & $155\ \rm{MeV}$ & $39\ \rm{MeV}$ & -- \\
\hline\hline
\end{tabular}
\end{center}
\label{table2}
\end{table*}

\section{Proton acceleration by petawatt-picosecond laser}
The results given in Table \ref{table1} demonstrate weak dependence
of energy ratio on the laser pulse duration, thus it is possible to
extend the qualitative model to estimate the proton energy in
picosecond laser solid interactions. However, in picosecond laser
plasma interactions, a significant amount of preplasmas are usually
generated by the irradiated prepulse prior to the arrival of the
main pulse. To check the validity of the model, we have repeated the
above simulations by considering two preplasma conditions.

In the first preplasma condition, a small preplasma is placed in front of the foil target, which
corresponds to a high-contrast laser pulse. The preplasma density
profile is $n_e=n_0\exp(x/l)$, where $l=1\ \rm{\mu m}$ and the total
preplasma length is $10\ \rm{\mu m}$. The flat foil target is
composed of copper plasma with electron number density 
$n_0=40n_c$. The foil thickness and width are $10\ \rm{\mu m}$ and
$34\ \rm{\mu m}$, respectively. A plastic layer as proton source, of
thickness $0.5\ \rm{\mu m}$, is attached to the foil rear surface.
Three simulations for laser intensities $1\times10^{19}\
\rm{W/cm^2}$, $5\times10^{19}\ \rm{W/cm^2}$, and $2\times10^{20}\
\rm{W/cm^2}$ are performed. The laser spot radius is $10\ \rm{\mu
m}$ and the laser pulse duration is $1\ \rm{ps}$. The laser
wavelength is $1.06\ \rm{\mu m}$. The simulation box $(x\times y)$
is $180\ \mathrm{\mu m}\times50\ \mathrm{\mu m}$ with the spatial
grids $6749\times1850$, respectively. The corresponding grid
length is 1 skin depth. The plastic layer is resolved by 19 spatial
grids. In each plastic target cell, there are 50 electrons, 25 ions,
and 1500 proton macroparticles.

The simulation results for the first preplasma condition are shown
in Table \ref{table2}. The proton energies in 2D simulations for the
laser intensities $1\times10^{19}\ \rm{W/cm^2}$, $5\times10^{19}\
\rm{W/cm^2}$, and $2\times10^{20}\ \rm{W/cm^2}$ are $36\ \rm{MeV}$,
$82\ \rm{MeV}$ and $155\ \rm{MeV}$, respectively. According to Eq.
\ref{eq2}, the energy ratio is 3.96 for laser with spot radius 
$10\ \rm{\mu m}$. After dividing the 2D simulation results by the
energy ratio, the corrected proton energies are $9\ \rm{MeV}$, $21\
\rm{MeV}$ and $36\ \rm{MeV}$, respectively. In view of the effects
of preplasma length on proton acceleration, we have also carried out
simulations with an $l=2\ \rm{\mu m}$ preplasma and total length
$20\ \rm{\mu m}$. The laser intensity is $5\times10^{19}\
\rm{W/cm^2}$. In this case, one can see from Fig. \ref{fig4}(b) that
the electron energy spectrum is similar to that of the $10\ \rm{\mu
m}$ preplasma. For the longer $20\ \rm{\mu m}$ preplasma both the
hot electron number and temperature are slightly higher than the
$10\ \rm{\mu m}$ preplasma, so that the TNSA proton energy ($85\
\rm{MeV}$ vs. $82\ \rm{MeV}$) is slightly higher than that of the
shorter preplasma case.

In order to characterize more realistic experimental conditions of
petawatt-picosecond laser facilities, three simulations are carried
out using the laser parameters from the current laser facility
SGII-U at Shanghai, China. The simulation and experimental results
are also given in Table \ref{table2}. The laser intensity in each
simulation is $5.0\times10^{19}\ \rm{W/cm^2}$, $5.8\times10^{19}\
\rm{W/cm^2}$, and $7.2\times10^{19}\ \rm{W/cm^2}$, respectively. The
laser spatial profile is Gaussian, the laser duration is $1\
\rm{ps}$, the laser spot radius is $10\ \rm{\mu m}$, and the laser
wavelength is $1.06\ \rm{\mu m}$. A preplasma with $l=8\ \rm{\mu m}$
is assumed in the simulations, and the total preplasma length is
$80\ \rm{\mu m}$. The width of the foil target is $50\ \rm{\mu m}$.
The thickness of the foil target is $10\ \rm{\mu m}$ and $15\
\rm{\mu m}$ in the two cases. The target rear surface is coated with
a plastic layer with thickness of $0.5\ \rm{\mu m}$. The simulation
box is  $(x, y) = (290, 60)\mathrm{\mu m}$ with a spatial grid of
($10872, 2250$). The corresponding grid length is 1 skin depth. The
other simulation parameters are same as our picosecond laser
simulations. For simulating such petawatt-picosecond laser
experiments, a full 3D simulation is far beyond our computational
resources, so that the 2D to 3D MPE ratio introduced here is useful.

\begin{figure}
\includegraphics[width=12cm]{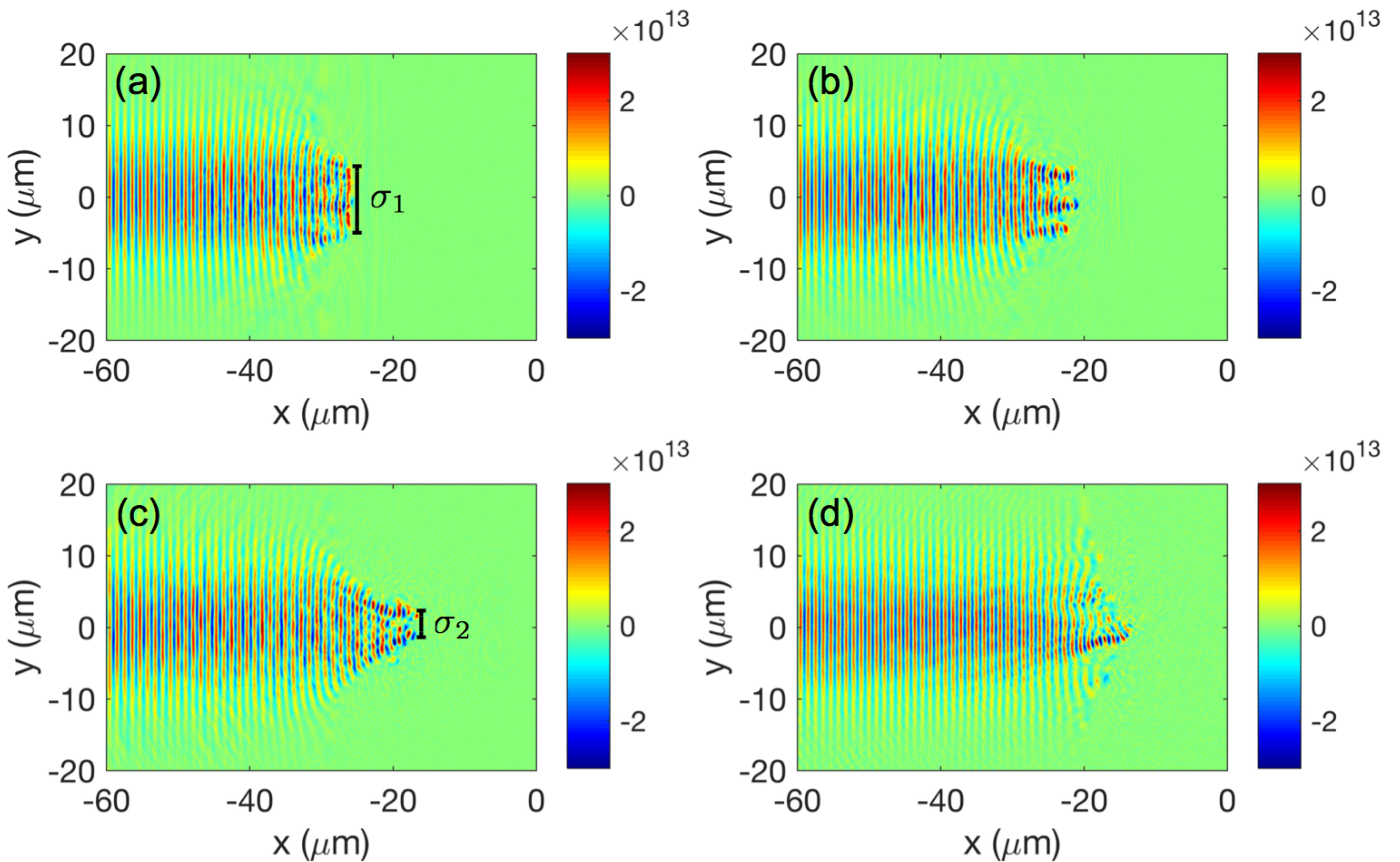}
\caption{\label{fig_pre} (Color online.) Distribution of the laser
electric field $E_y$ in unit of $\rm{V/m}$. (a) Snapshot at
$t=60T_0$, which is the time before the laser filamentation.
The laser spot radius at this time is about $\sigma_1\approx5\
\rm{\mu m}$. (b) Snapshot at  $t=70T_0$, when the laser
filamentation occurs. (c) Snapshot at  $t=90T_0$, when the
laser filaments merge into one central beam. The laser beam spot
radius at this time is about $\sigma_2\approx2.5\ \rm{\mu m}$. (d)
Snapshot at  $t=270T_0$, which is the time close to $1\
\rm{ps}$.}
\end{figure}

In cases with long-scale preplasmas, the laser propagation in
preplasma may be strongly affected by the nonlinear instabilities,
such as laser self-focusing, filamentation and hosing instability.
These instabilities can change the laser spot in the preplasma. But
the laser spot used in calculation of the energy ratio given in Eq.
\ref{eq2} is the vacuum spot radius. For example, Fig. \ref{fig_pre}
shows the laser propagation in the preplasma for the laser intensity
$5\times10^{19}\ \rm{W/cm^2}$. We see in Fig. \ref{fig_pre}(a) that
the laser spot radius changes from $\sigma_0$ in vacuum to
$\sigma_1$, and in Fig. \ref{fig_pre}(b) that modulation of the
pulse front and filamentation of the laser occur. Fig.
\ref{fig_pre}(c) shows that the laser pulse breaks up into several
filaments,\cite{pukhov_film, huang} and these filaments finally
merge into one central beam at the later time. The laser spot radius
at this time is denoted by $\sigma_2$. It is shown in Fig.
\ref{fig_pre} that the laser spot radius of $\sigma_1$ is about $5\
\rm{\mu m}$ and $\sigma_2$ is about $2.5\ \rm{\mu m}$.
From Eq. \ref{eq2}, we find that the energy ratios for $\sigma_0$,
$\sigma_1$, and $\sigma_2$ are 3.96, 2.80, and 1.98, respectively.
The MPE from the 2D simulation is $95\ \rm{MeV}$. By dividing the
energy ratios, the corrected proton energies for $\sigma_0$,
$\sigma_1$, and $\sigma_2$ are $23.2\ \rm{MeV}$, $33.9\ \rm{MeV}$,
and $47.9\ \rm{MeV}$, respectively. We see that the result
corresponding to the vacuum radius $\sigma_0$ matches well with the
experimental result $23.8\ \rm{MeV}$, but the results corresponding
to the self-focused lasers clearly over-estimate the proton energy.

From Table \ref{table2}, the proton energies in 2D simulations for
the laser intensities $5\times10^{19}\ \rm{W/cm^2}$,
$5.8\times10^{19}\ \rm{W/cm^2}$, and $7.2\times10^{19}\ \rm{W/cm^2}$
are $95\ \rm{MeV}$, $107\ \rm{MeV}$, and $113\ \rm{MeV}$,
respectively. After dividing the 2D simulation results by the energy
ratio, the corrected proton energies are $23.8\ \rm{MeV}$, $26.9\ 
\rm{MeV}$, and $28.3\ \rm{MeV}$, respectively. The experiment results
are $23.2\ \rm{MeV}$, $24.8\ \rm{MeV}$, and $29.1\ \rm{MeV}$,
respectively. The corrected proton energy agrees fairly well with
the experiment result.

Our result shows that by using 2D simulations, one can still predict
the MPE in the picosecond laser-solid target experiments at
the SGII-U laser facility.
However, for picosecond laser pulses with lower contrast, larger
scale preplasma will be generated and the laser pulse can be
affected by the self-focusing, filamentation, and hosing
instabilities. The TNSA of protons can then be affected, so that our
empirical energy ratio may not be applicable.

\begin{figure*}
\includegraphics[width=15cm]{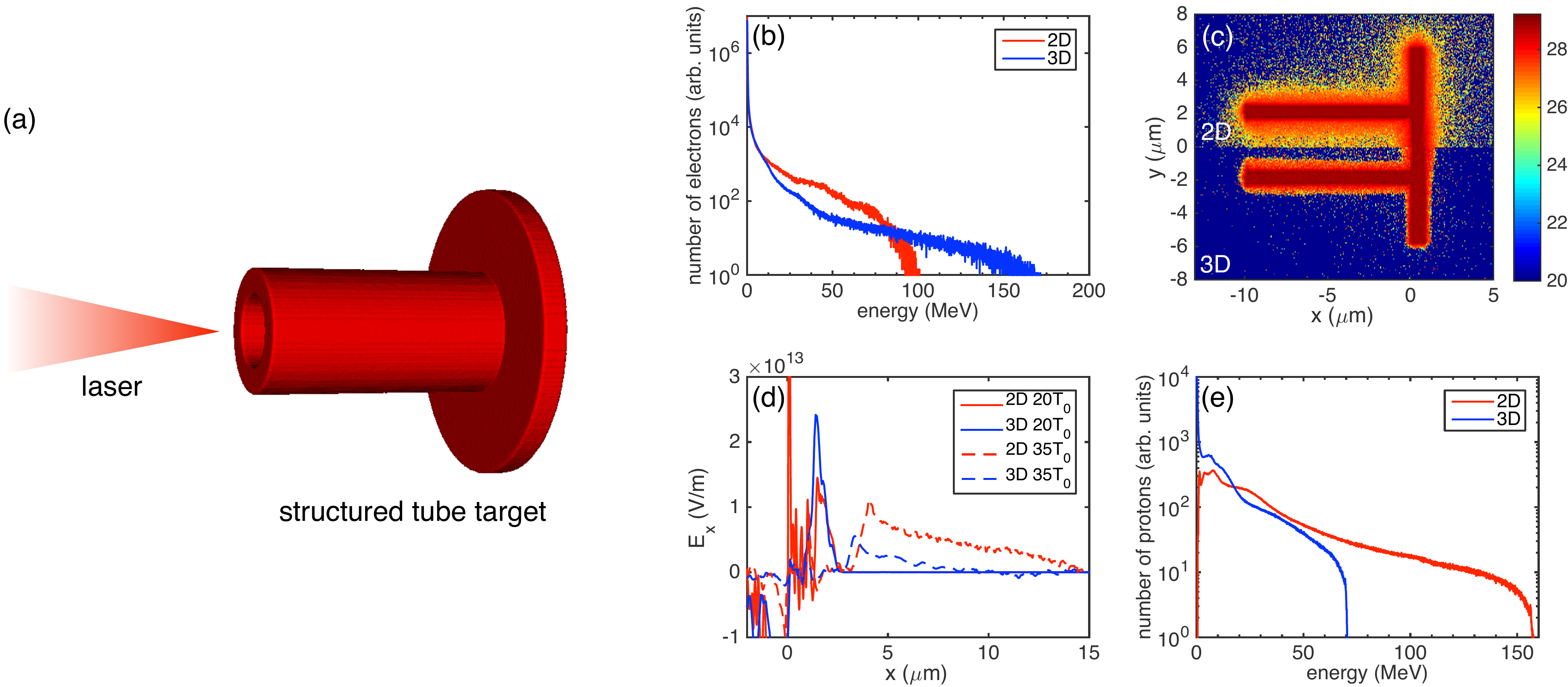}
\caption{\label{fig2} (Color online) Simulation results for the
structured tube target. (a) Target setup. (b) Electron spectrum at
$t=30{T_0}$. (c) Electron number density at  $t=100{T_0}$. (d)
Profile of $E_x$ at  $t=20{T_0}$ and $35{T_0}$, respectively. (e)
Proton spectrum at  $t=100{T_0}$.}
\end{figure*}

\begin{figure}
\includegraphics[width=11cm]{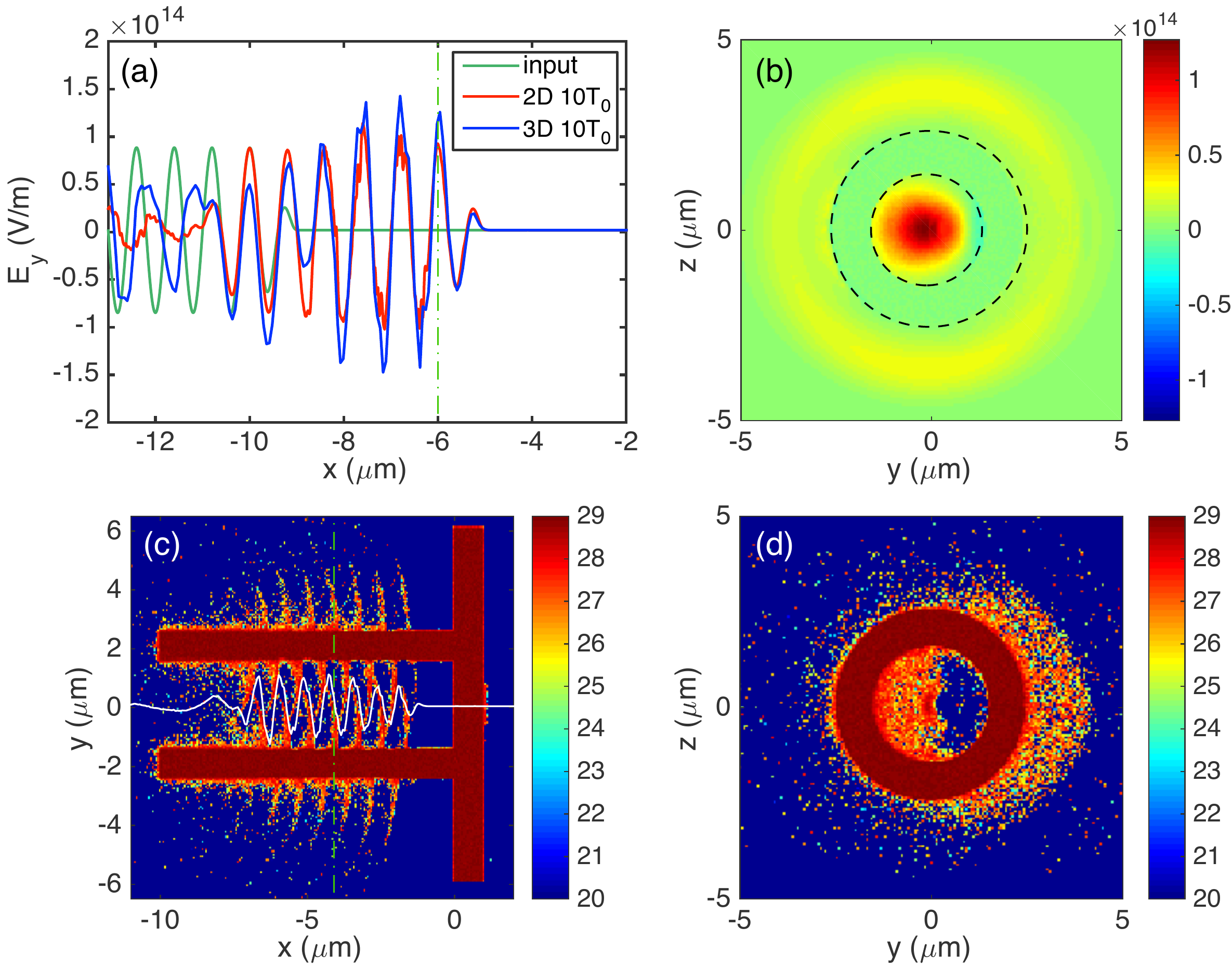}
\caption{\label{fig3} (Color online) (a) Electric field $E_y$ at
$t=10{T_0}$ for 2D and 3D simulation results, respectively. (b)
Transection of electric field $E_y$ in 3D simulation at $t=10{T_0}$
and position $x=-6\ \rm{\mu m}$. (c) Electron number density and
profile of the electric field $E_y$ (white line) in 3D simulation at
$t=15{T_0}$. (d) Transection of electron number density at
$t=15{T_0}$ and position $x=-4\ \rm{\mu m}$.}
\end{figure}

\begin{table*} \scriptsize
\caption{Comparison of maximum proton energy in 2D and 3D
simulations for the structured tube targets. 
The cone-tube target is the structured tube
target with an additional cone attached at the head of the plasma
tube.}
\begin{center}
\begin{tabular}{cccccc}
\hline\hline
target structure & laser intensity & spot and duration & proton energy (3D) & proton energy (2D) & energy ratio (2D/3D)\\
\hline
cone-tube target & $2\times10^{20}\ \rm{W/cm^2}$ & $4\ \rm{\mu m}$, $30\ \rm{fs}$ & $30\ \rm{MeV}$ & $78\ \rm{MeV}$ & 2.60 \\
cone-tube target & $5\times10^{20}\ \rm{W/cm^2}$ & $4\ \rm{\mu m}$, $30\ \rm{fs}$ & $62\ \rm{MeV}$ & $153\ \rm{MeV}$ & 2.46 \\
structured tube target & $1\times10^{21}\ \rm{W/cm^2}$ & $3\ \rm{\mu m}$, $20\ \rm{fs}$ & $71\ \rm{MeV}$ & $156\ \rm{MeV}$ & 2.19 \\
structured tube target & $5\times10^{19}\ \rm{W/cm^2}$ & $10\ \rm{\mu m}$, $1\ \rm{ps}$ & --  & $167\ \rm{MeV}$ & -- \\
\hline\hline
\end{tabular}
\end{center}
\label{table3}
\end{table*}

\section{Enhancement of proton energy using structured tube target}

In this section, multidimensional effects on laser interaction with
structured targets are investigated. Recently, many schemes have
been proposed to enhance the TNSA accelerated proton energy by using
the structured
targets.\cite{schwoerer,Toncian,Bartal,kluge,robinson,zhou-pro2,bqiao_pre,kdxiao_pro,jill,jiang,yilq,lecz,hulx}
The cone structure has been included in many target design, such as the slice-cone target\cite{zhengj}, the special designed target with two-stage acceleration\cite{liujl}, etc. 
In our simulation, we proposed to use a straight tube target, which composed of a
hollow cylinder plasma tube and a backside flat foil, as shown in
Fig. \ref{fig2} (a). 
Different from the cone target cases, in this scheme, the plasma tube acts as a waveguide. A periodic longitudinal electric field pattern is generated inside the tube and most of the electrons are effectively accelerated by this field. As a result, the proton energy is much higher than the normal foil cases\cite{xiao-aip}.

In the simulation, the inner radius and thickness of the plasma
tube are $1.5\ \rm{\mu m}$ and $1\ \rm{\mu m}$, respectively. The
thickness and width of the backside foil are $1\ \rm{\mu m}$ and
$12\ \rm{\mu m}$, respectively. Other simulation parameters are same
as the simulations given in Sec. II. In simulation, the Gaussian
laser irradiates from the left boundary of the simulation box and
injects into the hollow plasma tube. In the tube, the laser is
focused due to the optical confinement by the finite space inside
the hollow tube. During focusing, the laser intensity is increased
and the laser spot radius is reduced [see Figs. \ref{fig3} (a) and
\ref{fig3} (b)]. The laser focusing in 2D and 3D simulations are
compared. There is no $z$ direction in the 2D simulations, thus the
target in 2D is a planar object composed of two tube walls. The
injected laser is focused only in the $y$ direction by the upper and
bottom tube walls. However, in the 3D simulation, the tube focuses
the laser in the $y$ and $z$ directions simultaneously. A better
laser optical focusing is obtained by the spatial symmetric cylinder
tube [see Fig. \ref{fig3} (b)]. When the laser injects into the
tube, the laser $E_y$ field pulls the electrons out of the tube
walls. The pulled out electrons are then trapped and accelerated by
the focused laser field. These electrons are kept in the laser
accelerating phase for a long distance, comoving with the
propagating field in the form of a series of electron bunches [see
Figs. \ref{fig3} (c) and \ref{fig3} (d)]. The effective electron
acceleration finally results in a higher electron temperature
relative to the flat foil case in Sec. II [see Figs. \ref{fig1} (b)
and \ref{fig2} (b)].

Due to the stronger focusing of the laser field in the 3D
simulation, the electron maximum energy and high energy electron
temperature (for electrons with energies greater than $20\
\rm{MeV}$) in 3D simulation is higher than those in 2D [see Fig.
\ref{fig2}(b)]. Hence, at the beginning, the sheath field strength
in 3D is larger than that in 2D, which are
$E_x^{\rm{(3D)}}=2.5\times10^{13}\ \rm{V/m}$ in 3D and
$E_x^{\rm{(2D)}}=1.2\times10^{13}\ \rm{V/m}$ in 2D at $t=20{T_0}$.
However, in 3D, the electron divergence in the $z$ direction results
in lower trapping rate of the electrons pulled out by the laser
field. The total high-energy electron number in the 3D simulation is
less than that in 2D. [see Fig. \ref{fig2}(c)]. Moreover, in 3D
simulation the electron divergence at the target backside also
decreases the electron number density. Thus, at later times the
sheath field strength in the 3D simulation decays rapidly and
eventually lower than that in 2D. At $t=35{T_0}$, the sheath
field is $E_x^{\rm{(3D)}}=0.5\times10^{13}\ \rm{V/m}$ in 3D and
$E_x^{\rm{(2D)}}=1.0\times10^{13}\ \rm{V/m}$ in 2D. The proton
spectrum is shown in Fig. \ref{fig2}(e). The MPE in 3D is lower than
that in 2D, which are $71\ \rm{MeV}$ in 3D and $156\ \rm{MeV}$ in
2D, respectively. The ratio of the 2D to 3D MPE is 2.19, which
agrees well with the value 2.17 from Eq. \ref{eq2}. Compared with
the results given in Sec. II, the proton energy from the structured tube target
is higher than that from the foil target for both 2D and 3D, due to
the higher electron temperature in tube target. The simulation
results for different laser and target parameters are listed in
Table \ref{table3}. The cone-tube target is the structured tube
target with an additional cone attached at the head of the plasma
tube to increase the laser focusing \cite{xiao-aip}. It is shown
that the energy ratio for structured tube target agrees with the
qualitative model result given in Eq. \ref{eq2}.

To estimate the energy enhancement in petawatt-picosecond laser
cases, a 2D simulation is performed. The target is composed of a
plasma tube attached at the foil front surface with a small scale
preplasma. The inner radius and thickness of the plasma tube are $4\
\rm{\mu m}$ and $2\ \rm{\mu m}$, respectively. The length of the
plasma tube is $30\ \rm{\mu m}$. The thickness and width for the
backside foil are $2\ \rm{\mu m}$ and $34\ \rm{\mu m}$,
respectively. The target electron density is $n_0=40n_c$. The laser
intensity is $5\times10^{19}\ \rm{W/cm^2}$, and its duration is $1\
\rm{ps}$. The other simulation parameters are same as the
simulations with first preplasma condition in Sec. IV. It is found
that the MPE is $167\ \rm{MeV}$ in this 2D simulation. After
dividing by the energy ratio of 3.96, the corrected proton energy
would be $42\ \rm{MeV}$. The proton energy from the structured tube
target is about two times higher than the foil target.

\section{Summary}
In summary, multidimensional effects on TNSA of protons have been
investigated. Since the hot electron density and the induced sheath
field at the target rear surface decrease more rapidly in 3D than
that in 2D. The 2D simulations usually overestimate the MPE. Through
both 2D and 3D simulations, a qualitative scaling law is established
relating the MPEs obtained from the 2D and 3D simulations. It is
demonstrated that the MPE ratio depends strongly on the laser spot
size and displays weak dependence on the laser pulse durations,
which make it feasible to estimate the MPE in picosecond laser solid
interactions by only conducting the affordable 2D simulations. In
addition, it is also applicable to estimate the MPE in laser
structured target interactions by employing the energy ratio.

\begin{acknowledgments}
This work is supported by the National Key Program for S\&T Research
and Development, Grant No. 2016YFA0401100; the SSTDF, Grant No.
JCYJ20160308093947132; the National Natural Science Foundation of
China (NSFC), Grant Nos. 91230205, 11575031, and 11575298; the
National Basic Research 973 Project, Grant No. 2013CBA01500.  The
EPOCH code was developed under the UK EPSRC Grant Nos. EP/G054940/1,
EP/G055165/1, and EP/G056803/1. B.Q. acknowledges the support from
Thousand Young Talents Program of China.
We would like to thank that W. Wang presents us the experimental data 
done in the SGII-U laser facility in July 2017.
\end{acknowledgments}

{}

\end{document}